\documentclass[aps,prl,twocolumn,showpacs,superscriptaddress]{revtex4}
\usepackage[dvips]{graphicx}
\usepackage{subfigure}
\usepackage{amsmath}
\usepackage{amssymb}
\usepackage{times}

\begin{document}

\title{Modification of the Turbulent Energy Cascade by Polymer Additives}

\author{Nicholas T.~Ouellette}
\altaffiliation{Present address: Department of Physics, Haverford College, Haverford, PA 19041, USA}
\affiliation{International Collaboration for Turbulence Research}
\affiliation{Max Planck Institute for Dynamics and Self-Organization, D-37077 G\"{o}ttingen, Germany}

\author{Haitao Xu}
\affiliation{International Collaboration for Turbulence Research}
\affiliation{Max Planck Institute for Dynamics and Self-Organization, D-37077 G\"{o}ttingen, Germany}

\author{Eberhard Bodenschatz}
\email[Email: ]{eberhard.bodenschatz@ds.mpg.de}
\affiliation{International Collaboration for Turbulence Research}
\affiliation{Max Planck Institute for Dynamics and Self-Organization, D-37077 G\"{o}ttingen, Germany}
\affiliation{Laboratory of Atomic and Solid State Physics, Cornell University, Ithaca, NY 14853, USA}
\affiliation{Sibley School of Mechanical and Aerospace Engineering, Cornell University, Ithaca, NY 14853, USA}
\affiliation{Institute for Nonlinear Dynamics, Universit\"{a}t G\"{o}ttingen, D-37073 G\"{o}ttingen, Germany}

\date{\today}

\begin{abstract}
By tracking small particles in the bulk of an intensely turbulent flow, we show
that even a very small concentration of long-chain polymers disrupts the usual
turbulent energy cascade. The polymers affect scales much larger than their
physical size, from the dissipation range to the inertial range. The effect
depends strongly on the polymer concentration. While the dissipative-scale
statistics change continuously as the polymer concentration is increased, the
inertial-range energy transfer rate is only altered by the polymer additives
when the concentration is above a threshold (approximately 5 parts per million
by weight for the polymer we used).
\end{abstract}

\maketitle

Minute amounts of long-chain flexible polymers added to a fluid can strongly
modify flow properties. In a turbulent wall-bounded flow, for example, they
lead to the extraordinary phenomenon of drag reduction
\cite{toms:1948,virk:1967,lumley:1973,sreeni:2000,lvov:2004}.  These effects
may be qualitatively attributed to the stretching of polymer molecules by the
velocity gradients in the flow. Although progress has recently been made in
understanding drag reduction at a wall, comparatively little is known about the
action of polymers far from the boundaries of a turbulent
flow~\cite{liberzon:2005,liberzon:2006,crawford:2002,crawford:2004}. 

Fluid turbulence is inherently far from equilibrium: energy continuously passes
into and out of the system. In a Newtonian fluid like water, where molecular
viscosity provides the sole mechanism for the dissipation of energy, the
classical Richardson-Kolmogorov cascade hypothesis \cite{K41} states that
energy is injected into the flow at large length and time scales, transferred
to smaller and smaller scales without loss, and finally dissipated at the
smallest scales where viscosity acts.  The rates of energy \emph{injection}
$\epsilon_I$, energy \emph{transfer} $\epsilon_T$, and energy
\emph{dissipation} $\epsilon_D$, are therefore equivalent in Newtonian fluid
turbulence. Long-chain polymer molecules, which tend to coil up like balls of
thread in their equilibrium state, can be stretched by the straining of the
fluid flow to many times their equilibrium length and store elastic energy.  In
a turbulent flow, the polymer molecules will stretch and recoil in the
fluctuating flow field, and this process will dissipate kinetic energy due to
interactions between the monomers of one polymer molecule and between the
polymers and the fluid. The addition of polymers to a turbulent flow therefore
provides a new route by which kinetic energy can flow out of the turbulent
phase. The previously unaddressed but fundamental question is then the
relationship between $\epsilon_I$, $\epsilon_T$, and $\epsilon_D$ for
turbulence in polymer solutions.

In this Letter, we show that a very small concentration of long-chain polymer
molecules strongly modifies the turbulent cascade of energy from large to small
scales.  Our results indicate that the effect of polymers on the energy cascade
can be divided into two regimes depending on polymer concentration. For small
concentrations, the energy transfer rate $\epsilon_T$ is unchanged and only the
viscous dissipation rate $\epsilon_D$ is reduced due to the additional
dissipation mechanism provided by the polymers. For large concentrations,
however, we find that not only $\epsilon_D$ but also the apparent energy
transfer rate $\epsilon_T$ decreases.  The observations cannot be explained
by current theories.

Newtonian turbulence is described by a single nondimensional parameter, the
Reynolds number, which compares the strength of inertial driving forces to
viscous damping forces. A large Reynolds number implies that a wide range of
length and time scales participate in the energy cascade, where the statistical
properties of the flow are expected to be universal and independent of the
driving mechanism. This universal regime is known as the inertial range. A
polymer solution is additionally characterized by both the polymer
concentration and the Weissenberg number Wi, which compares $\tau_p$, the
relaxation time of a single polymer, to the fastest flow time scale.  In
turbulent flows, it is defined as $\textrm{Wi} = \tau_p / \tau_\eta$, where
$\tau_\eta$ is the Kolmogorov time scale. When the Weissenberg number is less
than a critical value (of order unity), the polymer molecules are generally in
their coiled state and will be passively advected by the flow. When the
Weissenberg number is larger than this value, the polymers will be stretched by
the flow and may modify it~\cite{lumley:1973}.

Earlier experimental investigations of the interactions of bulk turbulence and
polymers have generally either forced the turbulence through a boundary layer
~\cite{mccomb:1977,tong:1992,bonn:1993,vandoorn:1999}, 
or have been performed at relatively low Reynolds number~\cite{liberzon:2005,liberzon:2006}, 
where the turbulence was not fully
developed and it was difficult to quantify the effect of polymers on
turbulence, particularly for inertial-range quantities. 
{In a water flow between counter-rotating disks with raised vanes, it was observed that the energy injection at the disks remained constant when adding polymers into the flow~\cite{cadot:1998}.}
A previous experiment with polymers in our apparatus, however, showed that the
acceleration statistics of the flow were strongly affected~\cite{crawford:2002,crawford:2004}.  
Numerical simulations of isotropic
turbulence with polymers have also been performed at small Reynolds number
using model equations~\cite{vincenzi:2007,deangelis:2005,davoudi:2006,perlekar:2006}, but
they are very difficult due both to the nature of the equations and to
numerical instabilities~\cite{vaithi:2006}. 

In this Letter, we report results from Lagrangian particle tracking experiments
conducted in a water flow between counter-rotating disks with vanes.  
Flow properties were measured by
tracking~\cite{ouellette:2006} the simultaneous motion of hundreds of nearly
neutrally buoyant 33 $\mu$m fluorescent polystyrene tracer particles, excited
by a high-power pulsed Nd:YAG laser delivering up to 90 W and recorded with
three Phantom v7.1 CMOS cameras from Vision Research, Inc.  The polymer used
was an $18\times 10^6$ a.m.u. molecular weight polyacrylamide (Polysciences
18522) with an equilibrium radius of gyration of 0.5 $\mu$m, a fully stretched
length of 77 $\mu$m, and a relaxation time of $\tau_p = 43$
ms~\cite{crawford:2004}. While our apparatus allows us to reach very high
Reynolds numbers~\cite{laporta:2001,bourgoin:2006,voth:2002,ouellette:2006c},
the size and flexibility of the polymer molecules makes them prone to tearing
in very intense turbulence; we therefore only consider Reynolds numbers where
our results are not affected by polymer
degradation~\cite{crawford:2002,crawford:2004}.  In our experiments, the
Weissenberg and Reynolds numbers are coupled: based on the smallest turbulent
time scale, the Weissenberg number ranges from Wi = 1.2 to 6.0. The
Taylor-microscale Reynolds number $R_\lambda \equiv \sqrt{15 u' L / \nu}$
ranges from 200 to 350, where $u'$ is the root-mean-square turbulent velocity,
$L$ is the largest length scale of the turbulence, and $\nu$ is the kinematic
viscosity. We varied the polymer concentration from 0 (pure water) to 20 parts
per million by weight (ppm).  Note that, in order to make a quantitative
comparison with Newtonian fluid turbulence, the Reynolds numbers we report for
experiments with polymer solutions are those measured in pure water before
polymers were added to the flow.

\begin{figure}
\begin{center}
\subfigure[]{
\includegraphics[width=3.2in]{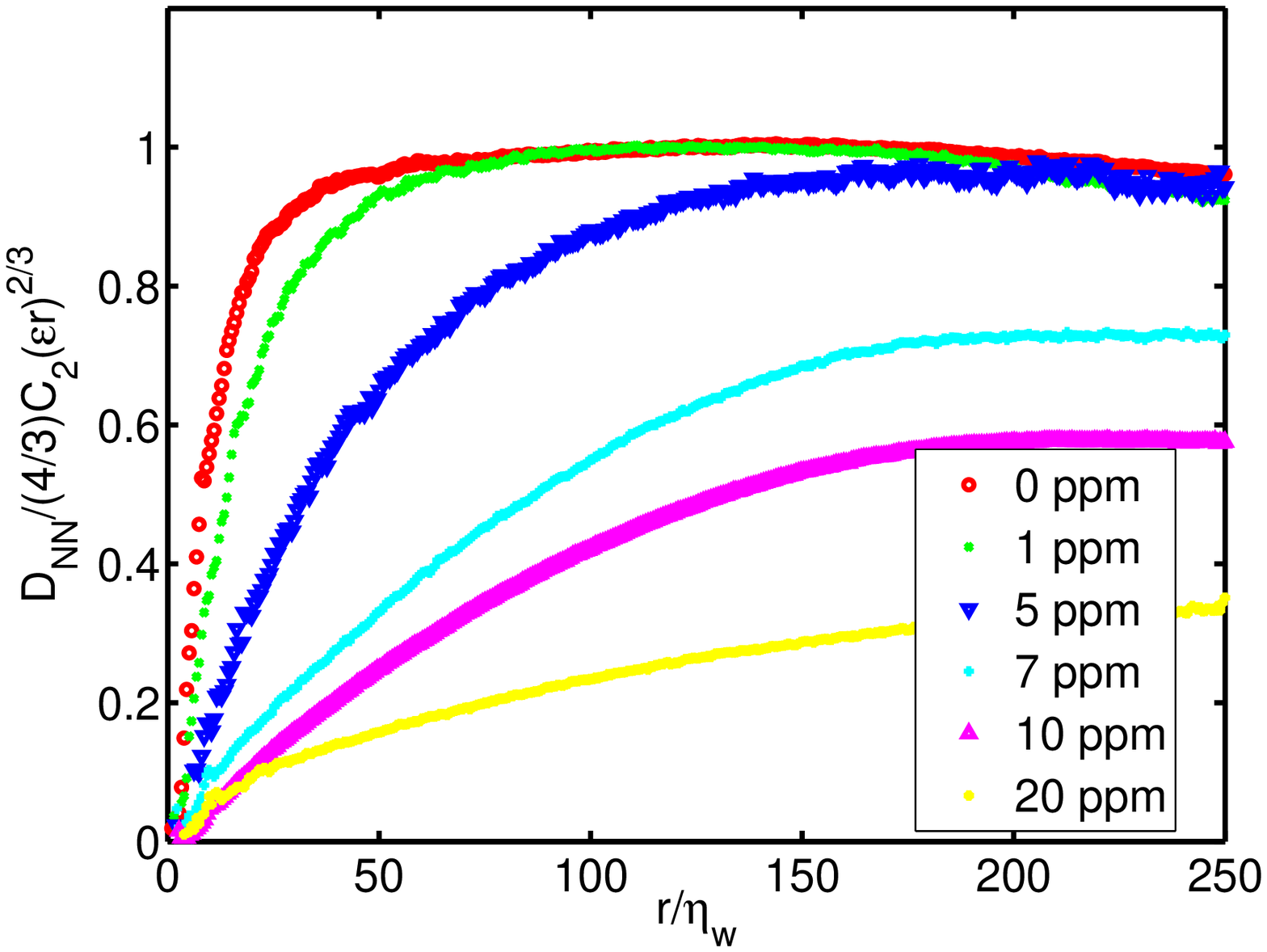}
}
\\
\subfigure[]{
\includegraphics[width=3.2in]{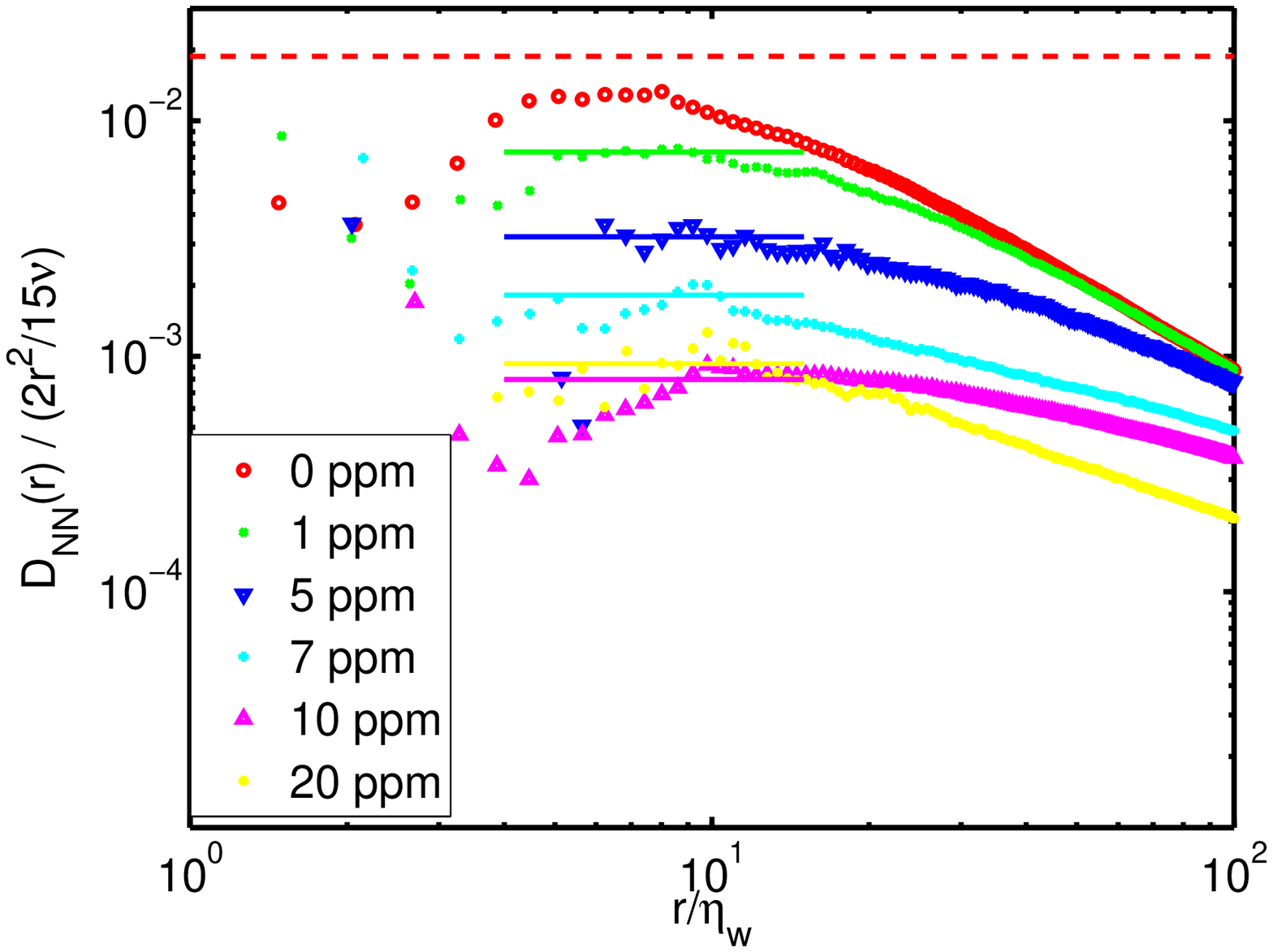}
}
\caption{(color online) The effect of polymer concentration on the Eulerian structure functions. The Reynolds number of the water flow (before adding polymers) is $R_\lambda = 350$ (corresponding to $\mathrm{Wi} = 6.0$ with polymers) and the Kolmogorov length scale is $\eta_w = 84\mu$m. The structure functions compensated by (a) the inertial-range scaling prediction and (b) the dissipation-range scaling prediction. The dashed line in (b) indicates $\epsilon_T(0)$ as measured in the water flow.}
\label{fig:DNN_compensate}
\end{center}
\end{figure}

To measure the modification of the turbulent energy cascade by the polymer
additives, we use the second-order transverse Eulerian structure function
$D_{NN}(r) \equiv \langle (\delta_r u)^2 \rangle $, which measures the
difference in velocity over a separation $\mathbf{r}$, as a probe of the
scale-by-scale properties of the cascade. For the transverse structure
function, velocities are measured orthogonal to $\mathbf{r}$. The analogous
longitudinal structure function $D_{LL}(r)$ is defined with the velocities
taken along $\mathbf{r}$. We show here only measurements of $D_{NN}(r)$; our
results are equivalent for $D_{LL}(r)$.  $D_{NN}(r)$ has three distinct scaling
regimes.  In the small-scale dissipation range ($r \ll \eta$), for isotropic
turbulence, 
\begin{equation} 
D_{NN}(r) = \frac{2\epsilon_D}{15 \nu} r^2, \quad
(r \ll \eta) .  \label{eq:DNN_smallr} 
\end{equation} 
At large scales ($r \gg L$), $D_{NN}(r)$ saturates at twice the velocity
variance.  At intermediate scales, in the so-called inertial range ($ \eta \ll
r \ll L$), the classical Kolmogorov theory~\cite{K41} predicts that
\begin{equation} 
D_{NN}(r) = \frac{4}{3} C_2 (\epsilon_T
r)^{2/3}, \quad (\eta \ll r \ll L), \label{eq:DNN_larger} 
\end{equation} 
where $C_2 = 2.13 \pm 0.22$ is a well-known universal constant determined from
previous experiments~\cite{sreeni:1995}.  We note again that in Newtonian
turbulence, the energy transfer rate $\epsilon_T$ and the energy dissipation
rate $\epsilon_D$ are the same.

We show in Fig.~\ref{fig:DNN_compensate} the measured $D_{NN}(r)$ for different
polymer concentrations at $R_\lambda = 350$, where $R_\lambda$ was measured
without polymers and the corresponding Kolmogorov length scale is $\eta_w = 84 \mu$m.  In Fig.~\ref{fig:DNN_compensate}(a), we plot $D_{NN}(r)$
compensated by the Kolmogorov scaling prediction (\textit{i.e.}, $\left[(3/4)
D_{NN}(r) / C_2\right]^{3/2} / (\epsilon_T(0) r)$, where $\epsilon_T(0)$ is the
energy transfer rate measured from the water data), so that a plateau at unity
indicates inertial-range scaling. We observe that at small concentrations
($\phi \leq 5$ ppm in our experiments), the shape of $D_{NN}(r)$ remains the
same: the curves return to the same plateau value in the inertial range, while
the extent of the apparent dissipation range increases with concentration. At
higher concentrations ($\phi \geq 7$ ppm), the apparent inertial-range plateau
is suppressed, indicating that the energy transfer rate has changed.  We also
measured the effect on the energy dissipation rate $\epsilon_D$ by plotting in
Fig.~\ref{fig:DNN_compensate}(b) $D_{NN}(r)$ compensated by the
dissipation-range scaling prediction (\textit{i.e.}, $\left[ 2\nu D_{NN}(r) /
15 r^2 \right]$). Due to the finite spatial resolution of our  measurement
system, the very small scales ($r \ll \eta$) are only partially resolved. 
 This effect is more pronounced for pure water case since for polymer solutions the small scales increases. Within experimental uncertainty, for water, $\epsilon_D(0) = \epsilon_T(0)$ as indicated by the dashed line in Fig.~\ref{fig:DNN_compensate}(b).
Nevertheless, we observe that the
effect of polymers on dissipation scales is smooth, without any sign of a transition.

\begin{figure}
\begin{center}
\includegraphics[width=3.2in]{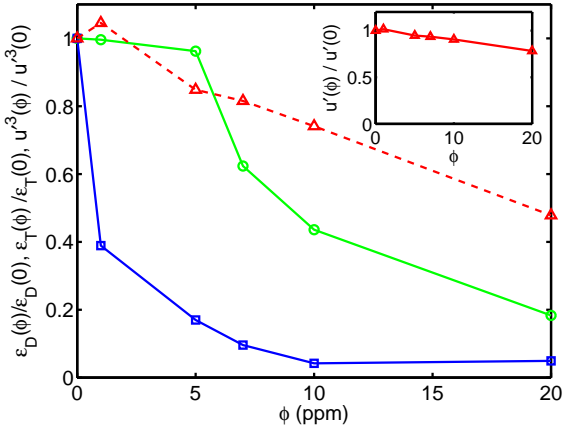}
\caption{(color online) Concentration effects on the energy transfer rate $\epsilon_T$ and energy dissipation rate $\epsilon_D$ at fixed Reynolds number ($R_\lambda = 350$; $\mathrm{Wi} = 6.0$). While $\epsilon_D$ ({\Large $\circ$}) decreases smoothly as the polymer concentration increases, the energy transfer rate $\epsilon_T$ ($\Box$) changes only when the concentration is above 5 ppm. $u'^3$ ($\triangle$) is shown as a surrogate for the energy injection rate $\epsilon_I$. The inset shows the change of $u'$ with concentration. The slow decrease of $u'$ is most likely due to the reduction of the forcing efficiency in the boundary layer of the propeller by drag reduction.}
\label{fig:eps_phi}
\end{center}
\end{figure}

To quantify the effect on the energy cascade, at each polymer concentration we
measured $\epsilon_T$ using Eq.~\ref{eq:DNN_smallr} and $\epsilon_D$ using
Eq.~\ref{eq:DNN_larger}, as shown in Fig.~\ref{fig:eps_phi}.  It is clear that
$\epsilon_T$ remains approximately unchanged for small concentrations, but
drops suddenly when the concentration is above 5 ppm. The energy dissipation
rate $\epsilon_D$, on the other hand, decreases smoothly with concentration.
We cannot measure the energy injection rate $\epsilon_I$ directly in our
current apparatus. Since $\epsilon_I \sim u'^3/L$, however, measurements of the
root-mean-square turbulent velocity can serve as a surrogate, provided that the
integral scale $L$ remains constant. The measured $u'$ decreases slowly with
concentration, as shown in Fig.~\ref{fig:eps_phi}. This nearly linear decay
cannot account for the change of $\epsilon_T$.

The physical basis for the observed transition at a concentration of roughly 5
ppm remains unclear. It is possible that it may result from polymer-polymer
interactions that occur only above the so-called ``overlap'' concentration.
For the polymer used in our experiments, the overlap concentration based on the
maximum extension length is $\sim 10^{-4}$ ppm, while it is approximately $200$
ppm if based on the radius of gyration. Neither of these estimates coincides
with the critical concentration observed in our experiments.

\begin{figure*}
\begin{center}
\includegraphics[width=2.3in]{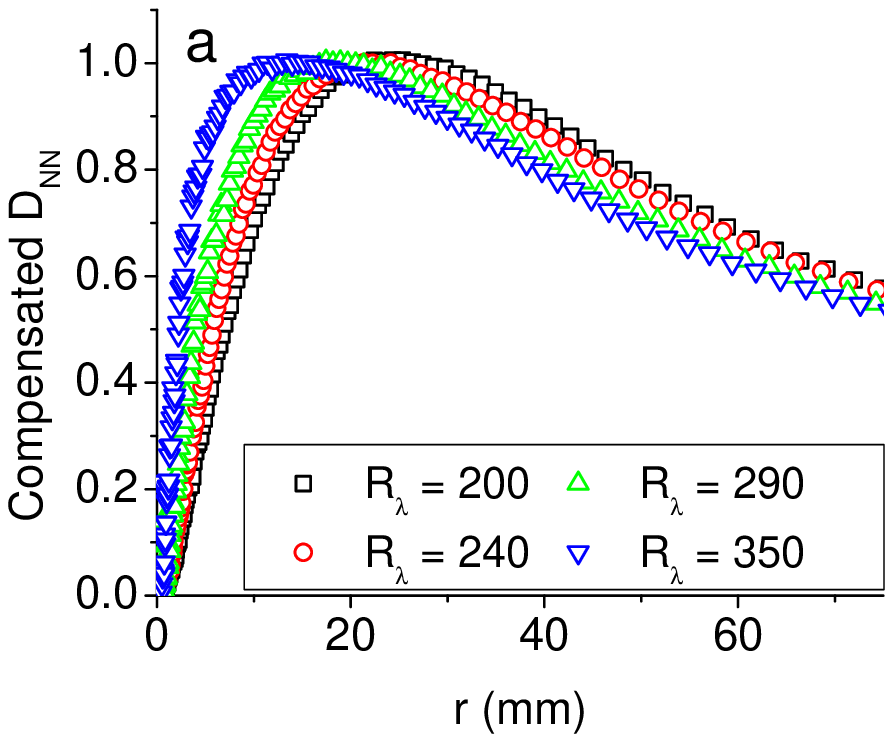}
\includegraphics[width=2.3in]{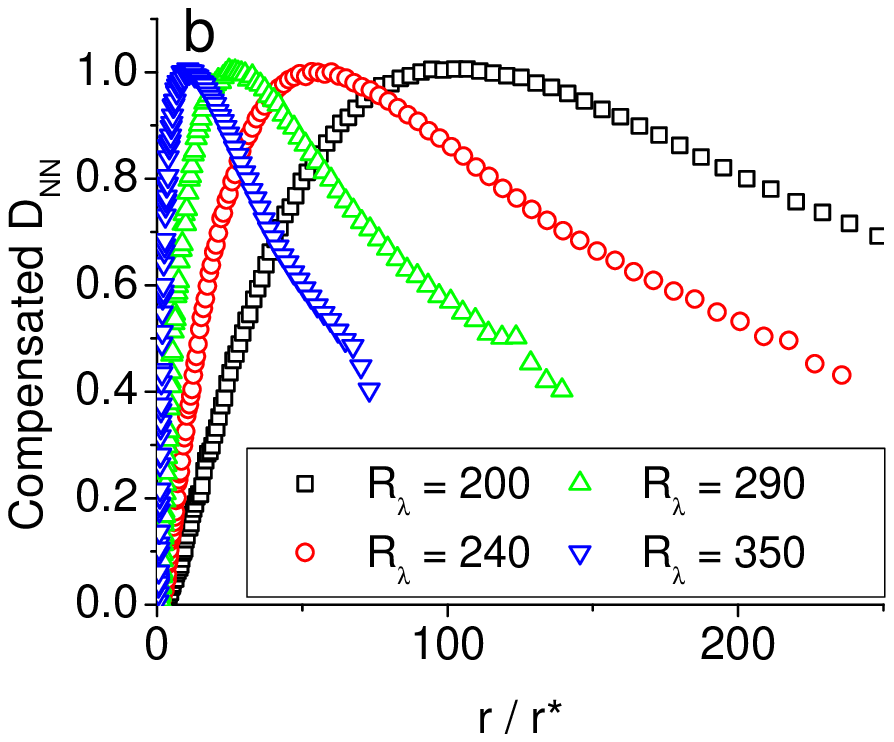}
\includegraphics[width=2.3in]{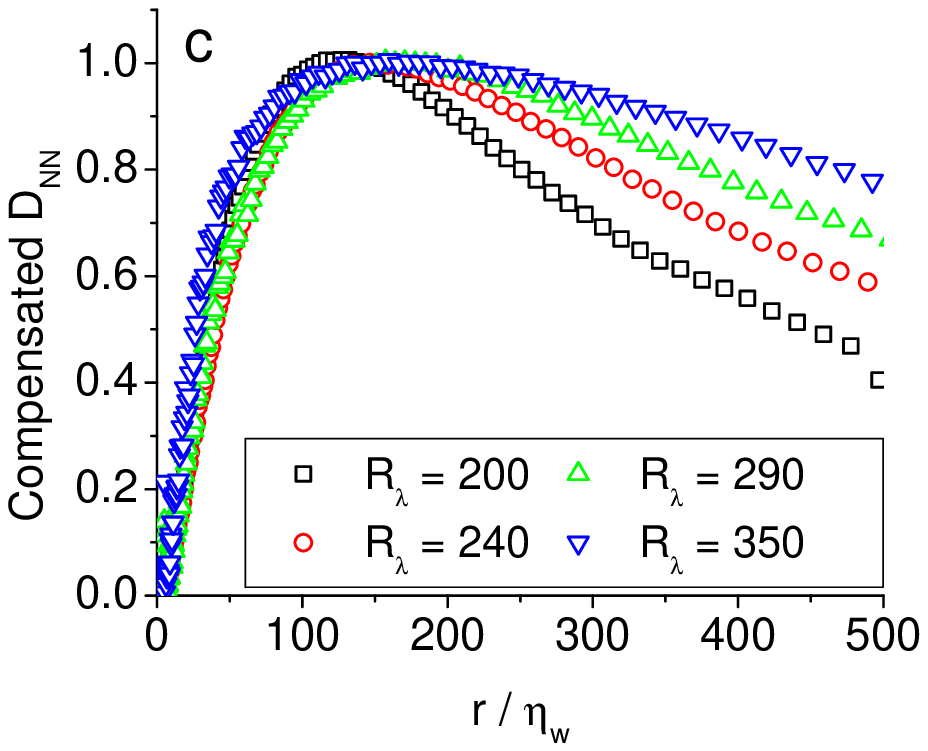}
\caption{(color online) Reynolds number effects at fixed concentration. Compensated Eulerian structure functions are shown for four Reynolds numbers in a 5 ppm polymer solution. (a) The length scale of the polymer effects changes with Reynolds number. (b) Scaling by the Lumley scale $r^*$ does not collapse the data for the different Reynolds numbers. (c) The small-scale data collapse when $r$ is scaled by $\eta$, the Kolmogorov length scale determined from water data.}
\label{fig:DNN_Re}
\end{center}
\end{figure*}

In his pioneering work in the 1970s, Lumley suggested that the length scale of
any polymer effect should be determined purely by the time scale on which the
polymer recoils \cite{lumley:1973}.  This (inertial-range) scale $r^* =
(\epsilon_T \tau_p^3)^{1/2}$ is the scale at which the local Weissenberg number
is unity, and at which we expect the polymers to begin to be stretched by the
flow.  In this theory, there is no concentration effect, and so it cannot fully
explain our experimental observations. It may, however, apply below the
critical concentration. Since the scale $r^*$ increases with Reynolds number,
we tested Lumley's hypothesis by varying the Reynolds number while keeping the
concentration fixed at 5ppm, just below the concentration where we observe
changes in the energy transfer rate.  The change of $D_{NN}(r)$ with
$R_\lambda$ is shown in Fig.~\ref{fig:DNN_Re}(a). As shown in
Fig.~\ref{fig:DNN_Re}(b), however, scaling by $r^*$ does not collapse our data.
We find instead, as illustrated in Fig.~\ref{fig:DNN_Re}(c), that the naive
normalization of $r$ by the Kolmogorov length scale $\eta$ (measured from pure
water at the same Reynolds number) collapses all the curves. We note that
$\eta$ \emph{decreases} with Reynolds number in our experiments, while the
prediction of $r^*$ \emph{increases}. The behavior we observe is therefore
qualitatively different from drag reduction at a boundary, where Lumley's
theory does appear to apply~\cite{lvov:2004}.

In contrast to Lumley's argument, Tabor and de Gennes suggested qualitatively
that while the polymers are affected by the flow at $r^*$, it is only at a
smaller scale $r^{**}$ that the flow is affected by the polymers
\cite{tabor:1986,degennes:1986}. In their framework, $r^{**}$ is determined by
balancing the turbulent kinetic energy at a given scale with the elastic energy
in the polymer phase. Each polymer molecule can store some elastic energy;
increasing the number of polymers therefore increases the energy in the polymer
phase. The Tabor--de Gennes picture thus allows for a concentration dependence
of the polymer effect.  There are, however, many undetermined parameters in
their qualitative theory, and further development is needed to make a
quantitative experimental test.

In summary, we investigated the effect of very small concentrations of
long-chain polymers on the dynamics of turbulence in the bulk of the flow.  We
observed a strong concentration dependence: the energy dissipation rate appears
to change for all polymer concentrations, while the inertial-range energy
transfer rate only changes above a critical concentration.  At a fixed
concentration below the critical concentration, we observed that the scales at
which the polymers affect the energy cascade \emph{decrease} with increasing
Reynolds number, in a manner similar to the dependence of the Kolmogorov scale
$\eta$ on Reynolds number.  Both the concentration effect and the Reynolds
number effect can not be explained by the theory (using Lumley's time
criterion) that has been used to explain the drag reduction phenomenon. Our
results suggest several challenges for future research. The qualitative change
in the polymer effect above the critical concentration must be explained, and
the exact ways in which the polymers change the three energy rates in
turbulence must be clarified. And finally, if Lumley's time criterion does not
hold in the bulk, a physical mechanism by which the polymers can affect scales
much larger than their size must be identified.

\begin{acknowledgments}
We are grateful to L.~Collins and D.~Vincenzi for helpful discussions over the course of this work, to M.~Gibert, W.~Pauls, and M.~Torralba for comments and suggestions for the manuscript, and to A.~Crawford for developing experimental protocols that give repeatable, robust results. This work was supported by the National Science Foundation under Grants No.~PHY-9988755 and No.~PHY-0216406 and by the Max Planck Society.
\end{acknowledgments}


\begin{thebibliography}{28}
\expandafter\ifx\csname natexlab\endcsname\relax\def\natexlab#1{#1}\fi
\expandafter\ifx\csname bibnamefont\endcsname\relax
  \def\bibnamefont#1{#1}\fi
\expandafter\ifx\csname bibfnamefont\endcsname\relax
  \def\bibfnamefont#1{#1}\fi
\expandafter\ifx\csname citenamefont\endcsname\relax
  \def\citenamefont#1{#1}\fi
\expandafter\ifx\csname url\endcsname\relax
  \def\url#1{\texttt{#1}}\fi
\expandafter\ifx\csname urlprefix\endcsname\relax\def\urlprefix{URL }\fi
\providecommand{\bibinfo}[2]{#2}
\providecommand{\eprint}[2][]{\url{#2}}

\bibitem[{\citenamefont{Toms}(1948)}]{toms:1948}
\bibinfo{author}{\bibfnamefont{B.~A.} \bibnamefont{Toms}},
  \bibinfo{journal}{Proc. 1$^{st}$ Intern. Rheol. Congr.}
  \textbf{\bibinfo{volume}{2}}, \bibinfo{pages}{135} (\bibinfo{year}{1948}).

\bibitem[{\citenamefont{Virk et~al.}(1967)\citenamefont{Virk, Merill, Mickley,
  Smith, and Mollo-Christensen}}]{virk:1967}
\bibinfo{author}{\bibfnamefont{P.~S.} \bibnamefont{Virk}},
  \bibinfo{author}{\bibfnamefont{E.~W.} \bibnamefont{Merill}},
  \bibinfo{author}{\bibfnamefont{H.~S.} \bibnamefont{Mickley}},
  \bibinfo{author}{\bibfnamefont{K.~A.} \bibnamefont{Smith}}, \bibnamefont{and}
  \bibinfo{author}{\bibfnamefont{E.~L.} \bibnamefont{Mollo-Christensen}},
  \bibinfo{journal}{J.~Fluid Mech.} \textbf{\bibinfo{volume}{30}},
  \bibinfo{pages}{305} (\bibinfo{year}{1967}).

\bibitem[{\citenamefont{Lumley}(1973)}]{lumley:1973}
\bibinfo{author}{\bibfnamefont{J.~L.} \bibnamefont{Lumley}},
  \bibinfo{journal}{J. Polymer Sci.: Macromolecular Reviews}
  \textbf{\bibinfo{volume}{7}}, \bibinfo{pages}{263} (\bibinfo{year}{1973}).

\bibitem[{\citenamefont{Sreenivasan and White}(2000)}]{sreeni:2000}
\bibinfo{author}{\bibfnamefont{K.~R.} \bibnamefont{Sreenivasan}}
  \bibnamefont{and} \bibinfo{author}{\bibfnamefont{C.~M.} \bibnamefont{White}},
  \bibinfo{journal}{J.~Fluid Mech.} \textbf{\bibinfo{volume}{409}},
  \bibinfo{pages}{149} (\bibinfo{year}{2000}).

\bibitem[{\citenamefont{L'vov et~al.}(2004)\citenamefont{L'vov, Pomyalov,
  Procaccia, and Tiberkevich}}]{lvov:2004}
\bibinfo{author}{\bibfnamefont{V.~S.} \bibnamefont{L'vov}},
  \bibinfo{author}{\bibfnamefont{A.}~\bibnamefont{Pomyalov}},
  \bibinfo{author}{\bibfnamefont{I.}~\bibnamefont{Procaccia}},
  \bibnamefont{and}
  \bibinfo{author}{\bibfnamefont{V.}~\bibnamefont{Tiberkevich}},
  \bibinfo{journal}{Phys.~Rev.~Lett.} \textbf{\bibinfo{volume}{92}},
  \bibinfo{pages}{244503} (\bibinfo{year}{2004}).

\bibitem[{\citenamefont{Liberzon et~al.}(2005)\citenamefont{Liberzon, Guala,
  L{\"u}thi, Kinzelbach, and Tsinober}}]{liberzon:2005}
\bibinfo{author}{\bibfnamefont{A.}~\bibnamefont{Liberzon}},
  \bibinfo{author}{\bibfnamefont{M.}~\bibnamefont{Guala}},
  \bibinfo{author}{\bibfnamefont{B.}~\bibnamefont{L{\"u}thi}},
  \bibinfo{author}{\bibfnamefont{W.}~\bibnamefont{Kinzelbach}},
  \bibnamefont{and} \bibinfo{author}{\bibfnamefont{A.}~\bibnamefont{Tsinober}},
  \bibinfo{journal}{Phys.~Fluids} \textbf{\bibinfo{volume}{17}},
  \bibinfo{pages}{031707} (\bibinfo{year}{2005}).

\bibitem[{\citenamefont{Liberzon et~al.}(2006)\citenamefont{Liberzon, Guala,
  Kinzelbach, and Tsinober}}]{liberzon:2006}
\bibinfo{author}{\bibfnamefont{A.}~\bibnamefont{Liberzon}},
  \bibinfo{author}{\bibfnamefont{M.}~\bibnamefont{Guala}},
  \bibinfo{author}{\bibfnamefont{W.}~\bibnamefont{Kinzelbach}},
  \bibnamefont{and} \bibinfo{author}{\bibfnamefont{A.}~\bibnamefont{Tsinober}},
  \bibinfo{journal}{Phys.~Fluids} \textbf{\bibinfo{volume}{18}},
  \bibinfo{pages}{125101} (\bibinfo{year}{2006}).

\bibitem[{\citenamefont{Crawford et~al.}(2002)\citenamefont{Crawford, La~Porta,
  Mordant, and Bodenschatz}}]{crawford:2002}
\bibinfo{author}{\bibfnamefont{A.~M.} \bibnamefont{Crawford}},
  \bibinfo{author}{\bibfnamefont{A.}~\bibnamefont{La~Porta}},
  \bibinfo{author}{\bibfnamefont{N.}~\bibnamefont{Mordant}}, \bibnamefont{and}
  \bibinfo{author}{\bibfnamefont{E.}~\bibnamefont{Bodenschatz}}, in
  \emph{\bibinfo{booktitle}{Advances in Turbulence IX}}, edited by
  \bibinfo{editor}{\bibfnamefont{I.~P.} \bibnamefont{Castro}},
  \bibinfo{editor}{\bibfnamefont{P.~E.} \bibnamefont{Hancock}},
  \bibnamefont{and} \bibinfo{editor}{\bibfnamefont{T.~G.} \bibnamefont{Thomas}}
  (\bibinfo{year}{2002}), p. \bibinfo{pages}{306}.

\bibitem[{\citenamefont{Crawford}(2004)}]{crawford:2004}
\bibinfo{author}{\bibfnamefont{A.~M.} \bibnamefont{Crawford}}, Ph.D. thesis,
  \bibinfo{school}{Cornell University} (\bibinfo{year}{2004}).

\bibitem[{\citenamefont{Kolmogorov}(1941)}]{K41}
\bibinfo{author}{\bibfnamefont{A.~N.} \bibnamefont{Kolmogorov}},
  \bibinfo{journal}{Dokl.~Akad.~Nauk SSSR} \textbf{\bibinfo{volume}{30}},
  \bibinfo{pages}{301} (\bibinfo{year}{1941}).

\bibitem[{\citenamefont{McComb et~al.}(1977)\citenamefont{McComb, Allan, and
  Greated}}]{mccomb:1977}
\bibinfo{author}{\bibfnamefont{W.~D.} \bibnamefont{McComb}},
  \bibinfo{author}{\bibfnamefont{J.}~\bibnamefont{Allan}}, \bibnamefont{and}
  \bibinfo{author}{\bibfnamefont{C.~A.} \bibnamefont{Greated}},
  \bibinfo{journal}{Phys.~Fluids} \textbf{\bibinfo{volume}{20}},
  \bibinfo{pages}{873} (\bibinfo{year}{1977}).

\bibitem[{\citenamefont{Tong et~al.}(1992)\citenamefont{Tong, Goldburg, and
  Huang}}]{tong:1992}
\bibinfo{author}{\bibfnamefont{P.}~\bibnamefont{Tong}},
  \bibinfo{author}{\bibfnamefont{W.~I.} \bibnamefont{Goldburg}},
  \bibnamefont{and} \bibinfo{author}{\bibfnamefont{J.~S.} \bibnamefont{Huang}},
  \bibinfo{journal}{Phys.~Rev.~A} \textbf{\bibinfo{volume}{45}},
  \bibinfo{pages}{7231} (\bibinfo{year}{1992}).

\bibitem[{\citenamefont{Bonn et~al.}(1993)\citenamefont{Bonn, Couder, van Dam,
  and Douady}}]{bonn:1993}
\bibinfo{author}{\bibfnamefont{D.}~\bibnamefont{Bonn}},
  \bibinfo{author}{\bibfnamefont{Y.}~\bibnamefont{Couder}},
  \bibinfo{author}{\bibfnamefont{P.~H.~J.} \bibnamefont{van Dam}},
  \bibnamefont{and} \bibinfo{author}{\bibfnamefont{S.}~\bibnamefont{Douady}},
  \bibinfo{journal}{Phys.~Rev.~E} \textbf{\bibinfo{volume}{47}},
  \bibinfo{pages}{R28} (\bibinfo{year}{1993}).

\bibitem[{\citenamefont{van Doorn et~al.}(1999)\citenamefont{van Doorn, White,
  and Sreenivasan}}]{vandoorn:1999}
\bibinfo{author}{\bibfnamefont{E.}~\bibnamefont{van Doorn}},
  \bibinfo{author}{\bibfnamefont{C.~M.} \bibnamefont{White}}, \bibnamefont{and}
  \bibinfo{author}{\bibfnamefont{K.~R.} \bibnamefont{Sreenivasan}},
  \bibinfo{journal}{Phys.~Fluids} \textbf{\bibinfo{volume}{11}},
  \bibinfo{pages}{2387} (\bibinfo{year}{1999}).

\bibitem[{\citenamefont{Cadot et~al.}(1998)\citenamefont{Cadot, Bonn, and
  Douady}}]{cadot:1998}
\bibinfo{author}{\bibfnamefont{O.}~\bibnamefont{Cadot}},
  \bibinfo{author}{\bibfnamefont{D.}~\bibnamefont{Bonn}}, \bibnamefont{and}
  \bibinfo{author}{\bibfnamefont{S.}~\bibnamefont{Douady}},
  \bibinfo{journal}{Phys.~Fluids} \textbf{\bibinfo{volume}{10}},
  \bibinfo{pages}{426} (\bibinfo{year}{1998}).

\bibitem[{\citenamefont{Vincenzi et~al.}(2007)\citenamefont{Vincenzi, Jin,
  Bodenschatz, and Collins}}]{vincenzi:2007}
\bibinfo{author}{\bibfnamefont{D.}~\bibnamefont{Vincenzi}},
  \bibinfo{author}{\bibfnamefont{S.}~\bibnamefont{Jin}},
  \bibinfo{author}{\bibfnamefont{E.}~\bibnamefont{Bodenschatz}},
  \bibnamefont{and} \bibinfo{author}{\bibfnamefont{L.~R.}
  \bibnamefont{Collins}}, \bibinfo{journal}{Phys.~Rev.~Lett.}
  \textbf{\bibinfo{volume}{98}}, \bibinfo{pages}{024503}
  (\bibinfo{year}{2007}).

\bibitem[{\citenamefont{De~Angelis et~al.}(2005)\citenamefont{De~Angelis,
  Casciola, Benzi, and Piva}}]{deangelis:2005}
\bibinfo{author}{\bibfnamefont{E.}~\bibnamefont{De~Angelis}},
  \bibinfo{author}{\bibfnamefont{C.~M.} \bibnamefont{Casciola}},
  \bibinfo{author}{\bibfnamefont{R.}~\bibnamefont{Benzi}}, \bibnamefont{and}
  \bibinfo{author}{\bibfnamefont{R.}~\bibnamefont{Piva}},
  \bibinfo{journal}{J.~Fluid Mech.} \textbf{\bibinfo{volume}{531}},
  \bibinfo{pages}{1} (\bibinfo{year}{2005}).

\bibitem[{\citenamefont{Davoudi and Schumacher}(2006)}]{davoudi:2006}
\bibinfo{author}{\bibfnamefont{J.}~\bibnamefont{Davoudi}} \bibnamefont{and}
  \bibinfo{author}{\bibfnamefont{J.}~\bibnamefont{Schumacher}},
  \bibinfo{journal}{Phys.~Fluids} \textbf{\bibinfo{volume}{18}},
  \bibinfo{pages}{025103} (\bibinfo{year}{2006}).

\bibitem[{\citenamefont{Perlekar et~al.}(2006)\citenamefont{Perlekar, Mitra,
  and Pandit}}]{perlekar:2006}
\bibinfo{author}{\bibfnamefont{P.}~\bibnamefont{Perlekar}},
  \bibinfo{author}{\bibfnamefont{D.}~\bibnamefont{Mitra}}, \bibnamefont{and}
  \bibinfo{author}{\bibfnamefont{R.}~\bibnamefont{Pandit}},
  \bibinfo{journal}{Phys.~Rev.~Lett.} \textbf{\bibinfo{volume}{97}},
  \bibinfo{pages}{264501} (\bibinfo{year}{2006}).

\bibitem[{\citenamefont{Vaithianathan et~al.}(2006)\citenamefont{Vaithianathan,
  Robert, Brasseur, and Collins}}]{vaithi:2006}
\bibinfo{author}{\bibfnamefont{T.}~\bibnamefont{Vaithianathan}},
  \bibinfo{author}{\bibfnamefont{A.}~\bibnamefont{Robert}},
  \bibinfo{author}{\bibfnamefont{J.~G.} \bibnamefont{Brasseur}},
  \bibnamefont{and} \bibinfo{author}{\bibfnamefont{L.~R.}
  \bibnamefont{Collins}}, \bibinfo{journal}{J. Non-Newtonian Fluid Mech.}
  \textbf{\bibinfo{volume}{140}}, \bibinfo{pages}{3} (\bibinfo{year}{2006}).

\bibitem[{\citenamefont{Ouellette
  et~al.}(2006{\natexlab{a}})\citenamefont{Ouellette, Xu, and
  Bodenschatz}}]{ouellette:2006}
\bibinfo{author}{\bibfnamefont{N.~T.} \bibnamefont{Ouellette}},
  \bibinfo{author}{\bibfnamefont{H.}~\bibnamefont{Xu}}, \bibnamefont{and}
  \bibinfo{author}{\bibfnamefont{E.}~\bibnamefont{Bodenschatz}},
  \bibinfo{journal}{Exp.~Fluids} \textbf{\bibinfo{volume}{40}},
  \bibinfo{pages}{301} (\bibinfo{year}{2006}{\natexlab{a}}).

\bibitem[{\citenamefont{La~Porta et~al.}(2001)\citenamefont{La~Porta, Voth,
  Crawford, Alexander, and Bodenschatz}}]{laporta:2001}
\bibinfo{author}{\bibfnamefont{A.}~\bibnamefont{La~Porta}},
  \bibinfo{author}{\bibfnamefont{G.~A.} \bibnamefont{Voth}},
  \bibinfo{author}{\bibfnamefont{A.~M.} \bibnamefont{Crawford}},
  \bibinfo{author}{\bibfnamefont{J.}~\bibnamefont{Alexander}},
  \bibnamefont{and}
  \bibinfo{author}{\bibfnamefont{E.}~\bibnamefont{Bodenschatz}},
  \bibinfo{journal}{Nature} \textbf{\bibinfo{volume}{409}},
  \bibinfo{pages}{1017} (\bibinfo{year}{2001}).

\bibitem[{\citenamefont{Bourgoin et~al.}(2006)\citenamefont{Bourgoin,
  Ouellette, Xu, Berg, and Bodenschatz}}]{bourgoin:2006}
\bibinfo{author}{\bibfnamefont{M.}~\bibnamefont{Bourgoin}},
  \bibinfo{author}{\bibfnamefont{N.~T.} \bibnamefont{Ouellette}},
  \bibinfo{author}{\bibfnamefont{H.}~\bibnamefont{Xu}},
  \bibinfo{author}{\bibfnamefont{J.}~\bibnamefont{Berg}}, \bibnamefont{and}
  \bibinfo{author}{\bibfnamefont{E.}~\bibnamefont{Bodenschatz}},
  \bibinfo{journal}{Science} \textbf{\bibinfo{volume}{311}},
  \bibinfo{pages}{835} (\bibinfo{year}{2006}).

\bibitem[{\citenamefont{Voth et~al.}(2002)\citenamefont{Voth, La~Porta,
  Crawford, Alexander, and Bodenschatz}}]{voth:2002}
\bibinfo{author}{\bibfnamefont{G.~A.} \bibnamefont{Voth}},
  \bibinfo{author}{\bibfnamefont{A.}~\bibnamefont{La~Porta}},
  \bibinfo{author}{\bibfnamefont{A.~M.} \bibnamefont{Crawford}},
  \bibinfo{author}{\bibfnamefont{J.}~\bibnamefont{Alexander}},
  \bibnamefont{and}
  \bibinfo{author}{\bibfnamefont{E.}~\bibnamefont{Bodenschatz}},
  \bibinfo{journal}{J.~Fluid Mech.} \textbf{\bibinfo{volume}{469}},
  \bibinfo{pages}{121} (\bibinfo{year}{2002}).

\bibitem[{\citenamefont{Ouellette
  et~al.}(2006{\natexlab{b}})\citenamefont{Ouellette, Xu, Bourgoin, and
  Bodenschatz}}]{ouellette:2006c}
\bibinfo{author}{\bibfnamefont{N.~T.} \bibnamefont{Ouellette}},
  \bibinfo{author}{\bibfnamefont{H.}~\bibnamefont{Xu}},
  \bibinfo{author}{\bibfnamefont{M.}~\bibnamefont{Bourgoin}}, \bibnamefont{and}
  \bibinfo{author}{\bibfnamefont{E.}~\bibnamefont{Bodenschatz}},
  \bibinfo{journal}{New J.~Phys.} \textbf{\bibinfo{volume}{8}},
  \bibinfo{pages}{109} (\bibinfo{year}{2006}{\natexlab{b}}).

\bibitem[{\citenamefont{Sreenivasan}(1995)}]{sreeni:1995}
\bibinfo{author}{\bibfnamefont{K.~R.} \bibnamefont{Sreenivasan}},
  \bibinfo{journal}{Phys. Fluids} \textbf{\bibinfo{volume}{7}},
  \bibinfo{pages}{2778} (\bibinfo{year}{1995}).

\bibitem[{\citenamefont{Tabor and {d}e Gennes}(1986)}]{tabor:1986}
\bibinfo{author}{\bibfnamefont{M.}~\bibnamefont{Tabor}} \bibnamefont{and}
  \bibinfo{author}{\bibfnamefont{P.~G.} \bibnamefont{{d}e Gennes}},
  \bibinfo{journal}{Europhys. Lett.} \textbf{\bibinfo{volume}{2}},
  \bibinfo{pages}{519} (\bibinfo{year}{1986}).

\bibitem[{\citenamefont{{d}e Gennes}(1986)}]{degennes:1986}
\bibinfo{author}{\bibfnamefont{P.~G.} \bibnamefont{{d}e Gennes}},
  \bibinfo{journal}{Physica A} \textbf{\bibinfo{volume}{140}},
  \bibinfo{pages}{9} (\bibinfo{year}{1986}).

\end{thebibliography}
\end{document}